\documentclass[prd,showpacs,twocolumn]{revtex4}
\usepackage{graphics}
\begin{document}
%
\title{Infrared cutoff dependence of the critical flavor number 
in three-dimensional QED}
\author{V.P. Gusynin$^{1*}$ and M. Reenders$^{2}$}
\affiliation{$^{1}$Department of Applied Mathematics, University of Western
Ontario, London, Ontario N6A 5B7, Canada\\
$^{2}$Department of Polymer Chemistry and Materials Science Center,\\
University of Groningen, Nijenborgh 4,
9747 AG Groningen, The Netherlands
}

\date{\today}
\begin{abstract}
We solve, analytically and numerically, a gap equation in parity
invariant QED$_3$ in the presence of an infrared cutoff $\mu$ and
derive an expression for the critical fermion number $N_c$ as a
function of $\mu$. We argue that this dependence of $N_c$ on the
infrared scale might solve the discrepancy between continuum
Schwinger-Dyson equations studies and lattice simulations of QED$_3$.
\end{abstract}
\pacs{11.10.Kk, 11.30.Qc, 12.20.Ds} 

\maketitle 

Parity invariant quantum electrodynamics
in 2+1 dimensions with $N$ flavors of four-component massless
fermions (QED$_3$) \cite{QED3} has been attracting a lot of 
interest for almost two decades. While it was often regarded
as a nice polygon for studying nonperturbative phenomena in gauge
field theories, such as dynamical mass generation, recently the
model has found applications in condensed matter physics, in
particular in high-$T_c$ superconductivity \cite{supercond}.

QED$_3$ is ultraviolet finite and has a built-in intrinsic mass scale
given by the dimensionful gauge coupling $e$, or $\alpha=e^2 N/8$,
which plays a role similar to the $\Lambda$ scale parameter in QCD.
In the leading order in $1/N$ expansion, it was found that at large
momenta ($p\gg \alpha$) the effective coupling between fermions and
gauge bosons vanishes (asymptotic freedom) whereas it has a finite
value or infrared stable fixed point at $p\ll \alpha$
\cite{apbokawij86}.  This behavior was shown to be robust against the
introduction of higher order $1/N$ corrections \cite{nash,guhare01}.
Studies of the gap equation for a fermion dynamical mass in the
leading order in the $1/N$ expansion have shown that massless QED$_3$
exhibits chiral symmetry breaking whenever the number of fermion
species $N$ is less than some critical value $N_c$, which is estimated
to be in the region $3<N_c<5$ ($N_c= 32/\pi^2\simeq3.2$ in the
simplest ladder approximation \cite{apnawij88}).  A renormalization
group analysis gives $3<N_c<4$ \cite{terao}.  Below such a critical
$N_c$, the $U(2N)$ flavor symmetry is broken down to $U(N)\times
U(N)$, the fermions acquire a dynamical mass, and $2 N^2$ Goldstone
bosons appear; for $N>N_c$ the particle spectrum of the model consists
of interacting massless fermions and a photon. In addition, it was
argued that the dynamical symmetry breaking phase transition at
$N=N_c$ is a conformal phase transition \cite{miya97,gumish98}.  The
last one is characterized by a scaling function for the dynamical
fermion mass with an essential singularity \cite{apnawij88,miya97}.

On the other hand, there is an argument due to Appelquist {\it et al.}
\cite{Appelquist} that $N_c\leq 3/2$. The argument is based on the 
inequality $f_{\rm IR}\leq f_{\rm UV}$ where $f$ is the thermodynamic 
free energy which is estimated in both infrared and ultraviolet regimes 
by counting massless degrees of freedom.

The above-described version of QED$_3$, the so-called parity invariant
noncompact version, was studied on a lattice \cite{lattice1}.  Recent
lattice simulations of QED$_3$ with $N\geq2$ have found no decisive
signal for chiral symmetry breaking \cite{lattice2}.  In particular,
for $N=2$, Ref.~\cite{lattice2} reports an upper bound for the chiral
condensate $\langle\bar\psi\psi\rangle$ to be of order $5\times
10^{-5}$ in units of $e^4$. We recall that for quenched, $N=0$,
QED$_3$ both numerical \cite{quenched_num} and analytical
\cite{dim_reg_qed} studies have shown that chiral symmetry is always
broken and numerically $\langle\bar\psi\psi\rangle\sim5\times10^{-3}$.
The above result of Ref.~\cite{lattice2} seemingly contradicts studies
based on Schwinger-Dyson (SD) equations which advocate dynamical
symmetry breaking for $N=2$ and seems to favor the estimate of
Appelquist {\it et al.} (we mention also that work in progress is
being done on lattice simulations of QED$_3$ with a single, $N=1$,
fermion flavor \cite{lattice3}).

One of the major problems in studying dynamical symmetry breaking 
using lattice simulations is to obtain control of finite size effects,
which play a nontrivial role
due to the presence of a massless photon \cite{lattice2}.
In terms of the intrinsic lattice spacing $a$, three dimensionless
length scales appear in lattice simulations:
the size of the lattice $L\sim 10-50$, 
the dimensionless lattice coupling constant $\beta\propto 1/(e^2 a)$, 
and the bare fermion mass $m_0a$.
In order to establish dynamical symmetry breaking the role of all these
scales in the problem should be well under control.
At present, lattice discretization appears to be well understood;
however, the finite size effects appear nontrivial and are most likely
the source of the discrepancy between lattice and continuum studies of 
QED$_3$ \cite{lattice2} .

In this paper, we present a possible explanation for the discrepancy
between recent lattice and continuum studies of QED$_3$ by studying
analytically and numerically the gap equation in the presence of the
infrared (IR) cutoff $\mu$, where $\mu$ is inversely related to the
size $L$ of a lattice. Since the characteristic ratio $e^2/\mu$ for
continuum QED$_3$ turns out to be proportional to the ratio $L/\beta$
in lattice simulations \cite{lattice2}, a comparison between two
approaches can be made. We will show that the presence of an IR cutoff
reduces the value of the critical number $N_c$ and derive the
relationship between $N_c$ and $\mu$.  Recently, a similar gap
equation, but with a massive photon, was studied numerically in
Refs.~\cite{cheng,pefr03} in connection with applications of QED$_3$
to high-$T_c$ superconductors (for an earlier numerical study in QED$_3$,
see Ref.~\cite{kona92}).

The gap equation of massless QED$_3$ with IR cutoff $\mu$ 
reads (compare with Eq.~(3.3) of Ref.\cite{apbokawij86})
\begin{eqnarray}
\Sigma(p)=\frac{\lambda\alpha}{2p} \int_\mu^\infty dk\,
\frac{k \Sigma(k)}{k^2+\Sigma^2(k)}\ln\frac{k+p+\alpha}{|k-p|+\alpha},
\label{nonlin:eq}
\end{eqnarray}
where $\lambda=8/N\pi^2$ and the Landau gauge is used.
Equation~(\ref{nonlin:eq}) is the simplest approximation to the SD
equation for the fermion self-energy which neglects corrections to the
fermion wave-function renormalization and vertex corrections.
Extensive studies showed that the nature of chiral symmetry breaking
which emerges from Eq.~(\ref{nonlin:eq}) with $\mu=0$ remains
qualitatively unchanged after including higher order corrections. In
what follows we solve Eq.~(\ref{nonlin:eq}) numerically but in order
to be able to treat it analytically one needs to make further
approximations. Because the integrand is damped at large momenta of
integration, the main contribution comes from the region with momenta
$k\ll\alpha$; thus, expanding the logarithm we come to the simplified
gap equation
\begin{eqnarray}
\Sigma(p)=\frac{\lambda}{p} \int_\mu^\alpha
dk\,\frac{k \Sigma(k)}{k^2+\Sigma^2(k)}{\rm min}(k,p).
\label{gapinteq}
\end{eqnarray}
The scale $\mu$ can be identified 
with the inverse size of the lattice in the temporal direction as 
$\mu\simeq \pi/La$ \cite{footnote}
since both scales $\mu$ and $\pi/La$ represent a minimal fermion 
momentum in continuum and lattice theories, respectively
[$\mu$ is related to the photon mass $m_a$
($\mu=m_a^2/\alpha$) in the model of Ref.~\cite{pefr03}]. 
Although this argument for the identification  of $\mu$ and $\pi/La$ 
(up to a factor of order 1) is certainly heuristic, it is 
not unreasonable.

On the other hand, the ultraviolet (UV) cutoff $\alpha$ in
Eq.~(\ref{gapinteq}) is in fact a physical scale which separates 
a nonperturbative dynamics at $k\ll\alpha$ 
from a perturbative one, with momenta much larger than $\alpha$, where
the dimensionless running coupling $\bar\alpha(k)=\alpha/(k+\alpha)$ 
\cite{apbokawij86} is weak. Neglecting those perturbative contributions 
in Eq.~(\ref{nonlin:eq}) results in an effective UV cutoff $\alpha$ in 
Eq.~(\ref{gapinteq}).

We shall solve the above equation (\ref{gapinteq}) using the
well-established linearized approximation when the momentum-dependent
dynamical mass in the denominator $k^2+\Sigma^2(k)$ is replaced by the
constant dynamical mass at the lower limit: $\Sigma^2(k)\rightarrow
\Sigma^2_0$, $\Sigma_0=\Sigma(\mu)$.  This approximation is known to
work well, especially near the phase transition point where
$\Sigma_0\ll\alpha$. As we show, the linearized equation
(\ref{gapinteq}) with an IR cutoff can be dealt with analytically and
an exact expression for $N_c$ as a function of $\alpha/\mu$ can be
derived, following the method developed in
Refs.~\cite{MGS,RNC,guhataya02}.

In the linearized approximation, Eq.~(\ref{gapinteq}) can be reduced 
to a hypergeometric differential equation
\begin{eqnarray}
u(1-u) \Sigma^{\prime\prime}(u)
+\frac{3}{2}(1-u)\Sigma^\prime(u)-\frac{\lambda}{4}\Sigma(u)=0,
\label{difeq2}
\end{eqnarray}
with the IR and UV boundary conditions 
\begin{eqnarray}
\left[p^2 \frac{d\Sigma}{dp}\right]_{p=\mu}=0,\qquad
\left[p\frac{d\Sigma}{dp}+\Sigma\right]_{p=\alpha}=0,\label{BCs}
\end{eqnarray}
where $u=-p^2/\Sigma_0^2$.
The general solution of Eq.~(\ref{difeq2}) can be written as
\begin{eqnarray}
\frac{\Sigma(p)}{\Sigma_0}=c_1 u_1(p)+c_2 u_2(p),
\end{eqnarray}
where we choose 
\begin{eqnarray}
&&u_1(p)=F\left(\frac{1+i\nu}{4},\frac{1-i\nu}{4};\frac{3}{2};
-\frac{p^2}{\Sigma_0^2}\right),\\
&&u_2(p)=\left(\frac{p}{\Sigma_0}\right)^{-(1+i\nu)/2}\nonumber\\
&\times&
F\left(-\frac{1-i\nu}{4},\frac{1+i\nu}{4};1+\frac{i\nu}{2};
-\frac{\Sigma_0^2}{p^2}\right)+\mbox{c.c.},
\end{eqnarray}
as a particular pair of independent solutions of Eq.~(\ref{difeq2}),
with $\nu=\sqrt{4\lambda-1}$, and where 
$F$ is the hypergeometric function.
If the infrared scale is set equal to zero ($\mu=0$), then
the IR boundary condition gives the constraint $c_2=0$,
since the function $u_2$ is too singular at $p=0$.
Then, the UV boundary condition fixes the relationship between the
dynamical mass $\Sigma_0$, $\nu$, and $\alpha$.
For nonzero $\mu$, the boundary conditions (\ref{BCs}) determine the
mass spectrum, giving rise to the equation
\begin{eqnarray}
f(\Sigma_0/\alpha,\mu/\alpha,\nu)=A_1 B_2-A_2 B_1=0, \label{gapeq2}
\end{eqnarray}
where 
\begin{eqnarray}
A_i\equiv \left[p\frac{d u_i}{dp}+u_i\right]_{p=\alpha},\quad
B_i\equiv\left[p^2 \frac{du_i}{dp}\right]_{p=\mu}.\label{A1A2B1B2def}
\end{eqnarray}
By making use of the formulas for differentiating the hypergeometric function 
\cite{erdelyi} the real functions $A_i$ and $B_i$ are expressed as
\begin{eqnarray}
A_1&=&F\left(\frac{1+i\nu}{4},\frac{1-i\nu}{4};\frac{1}{2};
-\frac{\alpha^2}{\Sigma_0^2}\right),\\
A_2&=&\left(\frac{\alpha}{\Sigma_0}\right)^{-(1+i\nu)/2}
\frac{(1-i\nu)}{2}\nonumber\\
&\times&
F\left(1-\frac{1-i\nu}{4},\frac{1+i\nu}{4};1+\frac{i\nu}{2};
-\frac{\Sigma_0^2}{\alpha^2}\right)+\mbox{c.c.}\nonumber\\
\end{eqnarray}
and
\begin{eqnarray}
B_1&=&-\mu \frac{\mu^2}{\Sigma_0^2} \frac{(1+\nu^2)}{12}\nonumber\\
&\times&F\left(1+\frac{1+i\nu}{4},1+\frac{1-i\nu}{4};\frac{5}{2};
-\frac{\mu^2}{\Sigma_0^2}\right),\label{B1}\\
B_2&=&-\mu\left(\frac{\mu}{\Sigma_0}\right)^{-(1+i\nu)/2}\frac{(1+i\nu)}{2}
\nonumber\\
&\times&F\left(1+\frac{1+i\nu}{4},-\frac{1-i\nu}{4};1+\frac{i\nu}{2};
-\frac{\Sigma_0^2}{\mu^2}\right)+\mbox{c.c.}\nonumber\\
&=&-\mu\left(\frac{\mu}{\Sigma_0}\right)^{-1/2}
\sqrt{1+\nu^2}\nonumber\\&\times&
\left[\mbox{Re}(F(\nu)) \cos\beta_2
+\mbox{Im}(F(\nu)) \sin\beta_2
\right],
\label{B2}
\end{eqnarray}
where $
\beta_2=({\nu}/{2})\ln({\mu}/{\Sigma_0})-\tan^{-1}\nu,$
and $F(\nu)$ is shorthand notation for the hypergeometric function: 
\begin{eqnarray}
F(\nu)=F\left(1+\frac{1+i\nu}{4},-\frac{1-i\nu}{4};1+\frac{i\nu}{2};
-\frac{\Sigma_0^2}{\mu^2}\right).\label{Fnudef}
\end{eqnarray}
Since we study the critical behavior, 
we can always assume that $\Sigma_0\ll \alpha$,
and therefore we can use the asymptotic expressions for
$A_1$ and $A_2$:
\begin{eqnarray}
A_1&\approx& \left(\frac{\alpha}{\Sigma_0}\right)^{-1/2}
\left[\frac{\Gamma(1/2)\Gamma(i\nu/2)}{\Gamma^2((1+i\nu)/4)} 
\left(\frac{\alpha}{\Sigma_0}\right)^{i\nu/2}+\mbox{c.c.}
\right]\nonumber\\
&=&\left(\frac{\alpha}{\Sigma_0}\right)^{-1/2}
|c(\nu)|\sqrt{1+\nu^2}
\cos\left(
\alpha_2+\theta\right)
\label{A1asym}
\end{eqnarray}
and
\begin{eqnarray}
A_2&\approx&\left(\frac{\alpha}{\Sigma_0}\right)^{-1/2}
\sqrt{1+\nu^2}\cos\alpha_2,
\label{A2asym}
\end{eqnarray}
where  
\begin{eqnarray}
&&\alpha_2=
\frac{\nu}{2}\ln \frac{\alpha}{\Sigma_0}+\tan^{-1}\nu,
\quad \theta=\arg c(\nu),\label{alpha2def}\nonumber
\\
&&c(\nu)=\frac{\Gamma(3/2)\Gamma(i\nu/2)}{
\Gamma((1+i\nu)/4)\Gamma((5+i\nu)/4)}.\nonumber
\end{eqnarray}
By making use of (2.10.2) of
\cite{erdelyi}, we can write $B_1$ in a form similar to $B_2$:
\begin{eqnarray}
B_1&=&-\mu\left(\frac{\mu}{\Sigma_0}\right)^{-(1+i\nu)/2}\frac{(1+i\nu)}{2}
c(-\nu)
\nonumber\\
&\times&F\left(1+\frac{1+i\nu}{4},-\frac{1-i\nu}{4};1+\frac{i\nu}{2};
-\frac{\Sigma_0^2}{\mu^2}\right)+\mbox{c.c.},
\nonumber\\
&=&
-\mu\left(\frac{\mu}{\Sigma_0}\right)^{-1/2}\sqrt{1+\nu^2}|c(\nu)|
\biggr[\mbox{Re}(F(\nu))\nonumber\\
&\times&\cos(\beta_2+\theta)
+\mbox{Im}(F(\nu)) \sin(\beta_2+\theta)
\biggr].
\label{B1simB2}
\end{eqnarray}
Finally, using the above expressions (\ref{B1})--(\ref{B1simB2}), the
gap equation (\ref{gapeq2}) can be expressed as
\begin{eqnarray}
f&=&-\mu|c|(1+\nu^2)
\left(\frac{\alpha \mu}{\Sigma_0^2}\right)^{-1/2}\sin\theta\bigg
[\mbox{Re}(F(\nu))\nonumber\\
&\times&
\sin(\beta_2-\alpha_2)
-\mbox{Im}(F(\nu))\cos(\beta_2-\alpha_2)
\bigg].
\label{gapeq_final}
\end{eqnarray}
One can convince oneself that for $\mu\ll\Sigma_0$ the last equation
is reduced to
\begin{equation}
\cos\left(\frac{\nu}{2}\ln\frac{\alpha}{\Sigma_0}+\theta+\tan^{-1}
\nu\right)=0,
\end{equation}
which gives the well-known result for the dynamical mass exhibiting
the essential singularity at $\nu\to0$:
\begin{equation}
\Sigma_0=\alpha\exp\left[-\frac{2(\pi n+\pi/2-\theta-\tan^{-1}\nu)}
{\nu}\right].
\label{mass_ess_sing}
\end{equation}
We recall that only the solution with $n=1$ corresponds to the stable 
ground state. Note also that for $N$ close to $N_c$ ($\nu\simeq0$) the 
dynamically generated mass $\Sigma_0$ is much less than the scale 
$\alpha$, providing a hierarchy of mass scales in the model under
consideration.

On the other hand, for $\Sigma_0\ll \mu$, we can use the power expansion
of $F(\nu)$ to get the equation for the dynamical mass $\Sigma_0$.
By expanding $F(\nu)$ of Eq.~(\ref{Fnudef}) for $\Sigma_0\ll \mu$, 
we obtain
\begin{eqnarray}
&&\mbox{Re}(F(\nu))=1+\frac{\Sigma_0^2}{\mu^2}
\rho(\nu^2)\cos\phi,\\
&&\mbox{Im}(F(\nu))=\frac{\Sigma_0^2}{\mu^2}
\rho(\nu^2)\sin\phi,
\end{eqnarray}
where 
\begin{eqnarray}
&&\rho(\nu^2)=\frac{1}{8}
\sqrt{\frac{(25+\nu^2)(1+\nu^2)}{4+\nu^2}},\\
&&\phi=-\tan^{-1}\frac{\nu(13+\nu^2)}{10-2\nu^2}.
\end{eqnarray}
Thus Eq.~(\ref{gapeq_final}) reduces to
\begin{eqnarray}
\frac{\Sigma_0^2}{\mu^2}=-\frac{1}{\rho(\nu^2)}
\frac{\sin\left(\frac{\nu}{2}\ln \frac{\alpha}{\mu}+2 \tan^{-1}\nu\right)}{
\sin\left(\frac{\nu}{2}\ln \frac{\alpha}{\mu}
+2\tan^{-1}\nu+\phi\right)}.\label{sigmaeq1}
\end{eqnarray}
Since, for $0<\nu<1$ $\phi$, is negative, we find that a nontrivial
or real solution for $\Sigma_0$ arises 
when $\nu$ exceeds the critical value $\nu_c$
determined by the equation 
\begin{eqnarray}
&&\frac{\nu_c}{2}\ln \frac{\alpha}{\mu}+2 \tan^{-1}\nu_c=\pi,\label{analytvc}
\end{eqnarray}
so that the $\sin$ in the numerator 
of Eq.~(\ref{sigmaeq1}) becomes negative (for $\nu>\nu_c$).
Note that the form of the gap equation (\ref{gapeq_final}) is different
in two regions $\mu\ll\Sigma_0$ and $\mu\gg\Sigma_0$: while in the
first one ($\mu\ll\Sigma_0$) we observe oscillations in the mass variable, 
in the second one ($\mu\gg\Sigma_0$) such oscillations disappear. This
is reflected also in the character of the mass dependence on the coupling
constant [compare Eqs.~(\ref{mass_ess_sing}) and (\ref{mfscallaw}) below].
In general, it can be shown that Eq.~(\ref{gapeq_final}) has $n$ nontrivial
solutions where the number $n$ is given by
\begin{eqnarray}
n=\left[\frac{\nu}{2\pi}\ln \frac{\alpha}{\mu}+\frac{2}{\pi}\tan^{-1}
\nu\right],
\end{eqnarray}
and the symbol $[C]$ denotes the integer part of the number $C$.
For small $\nu_c$, Eq.~(\ref{analytvc}) gives
\begin{eqnarray}
\nu_c=\frac{\pi}{2+({1}/{2})\ln(\alpha/\mu)}.
\end{eqnarray}
A similar functional dependence of $N_c$ was guessed in Ref.~\cite{pefr03}
by fitting their numerical study of the gap equation with a massive photon.

Near the critical point, we find from Eq.~(\ref{sigmaeq1}) the mean-field 
scaling law for $\Sigma_0$,
\begin{eqnarray}
\frac{\Sigma_0^2}{\mu^2}= \sigma (\nu-\nu_c), \label{mfscallaw}
\end{eqnarray}
where the positive ``nonuniversal'' 
constant of proportionality, $\sigma$, is
\begin{eqnarray}
\sigma=
\frac{1}{\rho(\nu_c^2)} 
\left[\frac{(\pi-2\tan^{-1}\nu_c)}{\nu_c}+\frac{2}{\nu_c^2+1}\right]
\frac{1}{\sin(\pi+\phi_c)},\nonumber
\end{eqnarray}
with $\phi_c$ the value of $\phi$ at $\nu=\nu_c$.

\begin{figure}
\rotatebox{-90}{
\resizebox*{0.7\columnwidth}{!}{\includegraphics{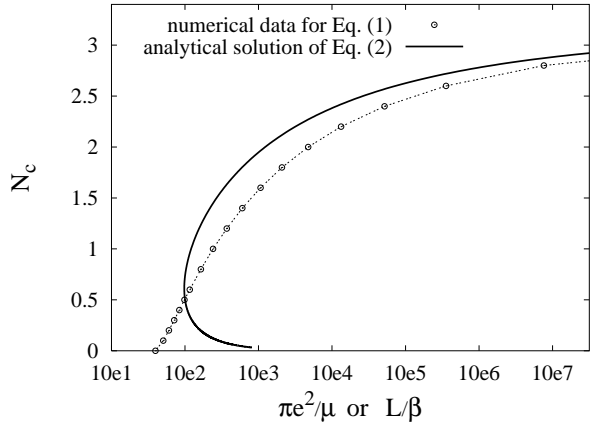}}}
\caption{The numerical solution of Eq.~(\ref{nonlin:eq}) and the 
analytical solution of Eq.~(\ref{gapinteq}) given by Eq.~(\ref{analytvc})  
for $N_c$ versus the ratio $\pi e^2/\mu$ or the lattice ratio
$L/\beta$.}
\label{fig_ncrit}
\end{figure}

We now compare the analytically obtained critical value of $\nu_c$
(Eq.~(\ref{analytvc})) or $N_c$ as a function of the ratio $\pi
e^2/\mu$ ($\alpha=N e^2/8$) with the numerical solution of the
integral equation (\ref{nonlin:eq}) which was obtained by using
numerical methods described in Ref.~\cite{guhare96}.  Basically,
starting with some initial guess for $\Sigma(p)$, the integral
equation is iterated using splines on a logarithmic momentum scale,
until sufficient precision is reached.  Both the analytical solution
for Eq.~(\ref{gapinteq}) and numerical solution for
Eq.~(\ref{nonlin:eq}) are depicted in Fig.~\ref{fig_ncrit}.  The
analytical solution of Eq.~(\ref{gapinteq}) agrees reasonable well
with the numerical solution of Eq.~(\ref{nonlin:eq}) for values of
$N\gtrsim 1$.  However, Eq.~(\ref{gapinteq}) (hence
Eq.~(\ref{analytvc})) provides a rather poor description of dynamical
symmetry breaking for small values of $N$ ( actually it does not allow
quenched limit in contrast to Eq.(\ref{nonlin:eq})).  The bending of
the curve for the analytic solution for $0\leq N \lesssim 1$ is due to
the linear dependence of the effective UV cutoff $\alpha$ on $N$.
Furthermore, the numerical solution shows the existence of a critical
ratio $\pi e^2/\mu \approx 40$ for $N=0$ (i.e., quenched QED$_3$) below
which no chiral symmetry breaking occurs.

From Fig.~\ref{fig_ncrit} it can be extracted that in order to find
chiral symmetry breaking for $N=2$ at least a ratio $\pi
e^2/\mu\approx 5\times10^3$ is required. Note that in order to draw 
decisive
conclusions on chiral symmetry breaking on a lattice, the volume of
the lattice $L^3$ must be large not just in dimensionless units but
also compared to any dynamically generated correlations in the system.
In particular, the physical volume $(L/\beta)^3$ is required to be
large \cite{lattice2}. Typical lattice simulations use at the most a
ratio $L/\beta\approx 50-100 $ and appear to lie outside the region
for dynamical symmetry breaking.  Indeed, by associating $La\simeq
\pi/\mu$ and $e^2\simeq 1/(\beta a)$, we have $L/\beta\simeq \pi
e^2/\mu$; thus, dynamical symmetry breaking near $N=2$ requires a ratio
$L/\beta\approx 5\times10^3$.  This explains the absence of any signs of
chiral symmetry breaking in lattice simulations with $N=2$. 
For $N=1$ the ratio is
$L/\beta\approx 240$ for the numerical solution of
Eq.~(\ref{nonlin:eq}); therefore, we believe that lattice simulations
with a single fermion flavor are crucial in establishing whether there
is a critical number $N_c$ in QED$_3$ and how this value depends on
the size $L$ of the lattice.  Moreover, as is evident from
Eq.~(\ref{mfscallaw}), the scaling near the critical point is expected 
to be of the mean-field type and such a scaling can also be checked 
using lattice simulations.

We are grateful to V.A.~Miransky for fruitful discussions and comments
and S.~Hands for valuable correspondence.  We thank A.H.~Hams for
help in the numerical programs.  V.P.G. acknowledges support from
the Natural Sciences and Engineering Research Council of Canada.  The
research of V.P.G. has been also supported in part by the
SCOPES-projects 7~IP~062607 and 7~UKPJ062150.00/1 of Swiss NSF and by
NSF Grant No. PHY-0070986.
%


\begin{thebibliography}{10}
\bibitem{footnote}Because of the antiperiodic boundary condition 
for fermions in the temporal direction, the coefficient 
is $\pi$ instead of $2\pi$.
\end{thebibliography}
\end{document}